\documentclass[manuscript,screen]{acmart}
\AtBeginDocument{%
  \providecommand\BibTeX{{%
    \normalfont B\kern-0.5em{\scshape i\kern-0.25em b}\kern-0.8em\TeX}}}

\setcopyright{acmcopyright}
\copyrightyear{2023}
\acmYear{2023}
\acmDOI{XXXXXXX.XXXXXXX}

\acmConference[CACM'23]{Submitted to Communications of the ACM}{October 10, 2023}{New York, NY}
\acmBooktitle{Communications of the ACM} 
\acmPrice{15.00}
\acmISBN{XXX-X-XXXX-XXXX-X/XX/XX}

\begin{document}

\title{Impact Factors for Computer Science Conferences}

\author{Carsten Eickhoff}
\email{c.eickhoff@acm.org}
\orcid{1234-5678-9012}
\affiliation{%
  \institution{University of T\"{u}bingen}
  \city{T\"{u}bingen}
  \country{Germany}
}

\renewcommand{\shortauthors}{Eickhoff}

\begin{abstract}
  An increasing number of CS researchers are employed in academic non-CS departments where publication output is measured in terms of journal impact factors. To foster recognition of publications in peer-reviewed CS conference proceedings, we analyzed more than 40,000 CS publications and computed journal impact factors for 88 top-ranking conferences across a representative range of fields, finding that some conferences have impact factors corresponding to those of high-ranking journals.
\end{abstract}





\maketitle

\section{Introduction}
Methods of Computer Science (CS) such as mobile sensing, artificial intelligence, or robotics, to name just a few, play an increasingly central role in other disciplines of research and practice. As a consequence, a rapidly increasing number of computer scientists are employed in academic non-CS departments, e.g.\ in the life  or engineering sciences. In stark difference to CS departments, many other disciplines evaluate scientific output of centers, labs, and individual researchers in terms of journal impact factors. Depending on local policies, these evaluations can affect tenure, promotion and resource allocation decisions. Since even highly selective conference proceedings do not receive ``official'' impact factors, the scientific output of CS researchers may appear artificially low. To make this form of publication output more directly comparable to established outlets such as journals, this article computes journal impact factors for the leading CS conferences.

There has been ample, justified criticism towards the use usefulness of impact factors as a measure of scientific productivity~\cite{hutchins2016relative, callaway2016beat,european2007ease}. This article does not challenge these observations. Much rather, it aims to put CS researchers in non-CS departments on equal footing with their journal-focused colleagues by providing the same widely used metric employed by scientometricians across departments.

\section{Methodology}
The selection of conferences for this study is based on the list of most impactful international CS venues according to \href{https://csrankings.org}{csrankings.org} (CSR) and the CORE Computing Research and Education ranking. CSR is a crowdsourced community project of CS faculty members tracking the publication output of CS departments and individual researchers in the leading conferences. CORE directly assigns letter grades to conferences. Both platforms are based on the contributions of hundreds of senior faculty-level volunteers around the globe. We included any CS conference listed on CSR (n=77), as well as those that received the highest ranking of A* on CORE (n=60). In total 88 conferences were included, with the majority satisfying both requirements (n=49).

For the included conferences, we computed two-year impact factors for 2022. This process requires collecting all articles published in a given conference in the two preceding years, i.e., in 2020 and 2021 and calculating the average number of citations each article received by any article published in the calendar year 2022.

We included all main-track articles (i.e., no extended abstracts, demonstrators, etc.) from the official proceedings and collected 2022 citation counts via Google Scholar. This methodology corresponds to the process employed by Clarivate, Thomson Reuters and others to compute journal impact factors. We adopted CSR's grouping of conferences into fields (AI, Systems, Theory, and Interdisciplinary) and research areas and manually incorporated those conferences only ranked by CORE into the same taxonomy. 

\section{Results}
Tables 1-4 show the result of our data analysis effort. In total, 44,746 papers were published across the 2020 and 2021 editions of the 88 considered conferences. For each conference, we computed the mean number of 2022 citations (the so-called \textit{impact factor}) as well as the more robust median. We additionally tracked selectiveness of conferences in terms of acceptance rates as well as the overall number of papers published per conference in 2020 and 2021. Overall, we see a wide spread of impact factors, ranging from 48.87 (ICLR) to 0.93 (EMSOFT), indicating both the brittleness of impact factors as a scientometric device, but also the different publication and citation standards in top-ranked venue of different fields and research areas.

\begin{table}
\caption{Scientometric overview of the field of AI, grouped by research areas, tracking impact factors, median number of citations per article, acceptance rates, and number of accepted articles $N$ per conference.}
\makebox[\textwidth][c]{
\begin{tabular}{lcccr|lcccr}
\toprule
\multicolumn{10}{l}{AI (IF 23.88)} \\
\midrule
Conf.\ & IF & median  & Acc.\ rate & $N$  & Conf.\ & IF & median  & Acc.\ rate & $N$ \\ \hline
\multicolumn{5}{l}{\textbf{Artificial Intelligence (IF 12.14)}} & \multicolumn{5}{l}{\textbf{Computer Vision (IF 34.70)}} \\
AAAI & 15.02 & 7 & 21.00\% & 3274 & CVPR & 44.08 & 23 & 22.90\% & 3126 \\
IJCAI & 7.87 & 5 & 13.15\% & 1178 & ECCV & 34.42 & 17 & 26.00\% & 1360 \\
AAMAS & 5.8 & 4 & 23.90\% & 337 & ICCV & 27.25 & 13 & 25.90\% & 1612 \\
ICAPS & 4.29 & 3 & 30.45\% & 89 & ACMMM & 9.9 & 6 & 27.90\% & 683  \\
KR & 3.96 & 2 & 30.00\% & 193 & \multicolumn{5}{l}{ }  \\
\midrule
\multicolumn{5}{l}{\textbf{Machine Learning (IF 27.11)}} & \multicolumn{5}{l}{\textbf{Natural Language Processing (IF 17.90)}} \\
ICLR    & 48.87 & 20 & 27.60\% & 1596 & ACL   & 25.66 & 12 & 24.95\% & 1351 \\
ICML    & 26.66 & 10 & 21.65\% & 2267 & NAACL & 16.49 & 8  & 29.20\% & 477 \\
NeurIPS & 23.27 & 9  & 22.90\% & 4236 & EMNLP & 15.68 & 8  & 25.05\% & 1599 \\
KDD     & 15.95 & 8  & 16.15\% & 456  & \multicolumn{5}{l}{ } \\
COLT     & 12.1 & 6 & 33.00\% & 255  & \multicolumn{5}{l}{ } \\
ISMAR     & 5.08 & 4 & 21.24\% & 125  & \multicolumn{5}{l}{ } \\
ICDM     & 4.59 & 2 & 9.85\% & 379  & \multicolumn{5}{l}{ } \\
\midrule
\multicolumn{5}{l}{\textbf{Web \& Information Retrieval (IF 12.46)}} & \multicolumn{5}{l}{ } \\
SIGIR & 19.41 & 9 & 23.75\% & 298 & \multicolumn{5}{l}{ } \\
WWW   & 12.99 & 7 & 19.90\% & 574 & \multicolumn{5}{l}{ } \\
\bottomrule
\end{tabular}%
}
\label{tab:addlabel}%
\end{table}%

\begin{table}
\caption{Scientometric overview of the field of Systems, grouped by research areas, tracking impact factors, median number of citations per article, acceptance rates, and number of accepted articles $N$ per conference.}
\makebox[\textwidth][c]{
\begin{tabular}{lcccr|lcccr}
\toprule
\multicolumn{10}{l}{Systems (IF 10.27)} \\
\midrule
Conf.\ & IF & median  & Acc.\ rate & $N$  & Conf.\ & IF & median  & Acc.\ rate & $N$ \\ \hline
\multicolumn{5}{l}{\textbf{Computer Architecture (IF 13.22)}} & \multicolumn{5}{l}{\textbf{Computer Networks (IF 9.79)}} \\
ASPLOS & 16.38 & 10 & 18.45\% & 197 & SIGCOMM & 16.94 & 13 & 22.21\% & 108 \\
HPCA & 14.07 & 8 & 21.50\% & 123 & NSDI & 14.04 & 10 & 17.03\% & 124 \\
ISCA & 12.02 & 7 & 18.30\% & 164 & INFOCOM & 7.79 & 5 & 19.40\% & 518 \\
MICRO & 10.2 & 6 & 22.00\% & 176 & IPSN & 4.75 & 4 & 22.67\% & 53 \\
\midrule
\multicolumn{5}{l}{\textbf{Computer Security (IF 15.98)}} & \multicolumn{5}{l}{\textbf{Databases (IF 9.05)}} \\
IEEE S\&P ``Oakland'' & 21.96 & 16 & 12.25\% & 219 & VLDB   & 10.36 & 6 & 24.46\% & 488 \\
USENIX Security       & 16.86 & 11 & 17.40\% & 402 & SIGMOD & 9.44 & 6 & 34.35\% & 332 \\
CCS                   & 12.85 & 7 & 19.60\% & 437 & ICDE & 7.25 & 4 & 22.70\% & 294 \\
\multicolumn{5}{l}{ }                               & PODS     & 3.91 & 2 & ?\% & 46  \\
\midrule
\multicolumn{5}{l}{\textbf{Design Automation (IF 4.00)}} & \multicolumn{5}{l}{\textbf{Embedded \& Real-Time Systems (IF 6.96)}} \\
ICCAD & 4.96 & 3 & 27.00\% & 271 & RTAS   & 5.88 & 4 & 25.50\% & 60 \\
DAC       & 3.45 & 1 & 23.10\% & 468 & RTSS & 4.33 & 2 & 22.20\% & 69 \\
\multicolumn{5}{l}{ }   & EMSOFT & 0.93 & 0 & 32.50\% & 28 \\
\midrule
\multicolumn{5}{l}{\textbf{High-Performance Computing (IF 7.26)}} & \multicolumn{5}{l}{\textbf{Mobile Computing (IF 11.57)}} \\
SC & 9.33 & 5 & 25.55\% & 205 & MobiCom   & 11.57 & 8 & 17.45\% & 122 \\
HPDC       & 7.0 & 3 & 21.44\% & 39 & SenSys & 10.28 & 8 & 19.30\% & 68 \\
PODC      & 5.44 & 3 & 25.34\% & 91   & MobiSys & 9.84 & 6 & 20.55\% & 70 \\
ICS       & 4.16 & 3 & 27.57\% & 81   & PERCOM & 6.7 & 4 & 15.00\% & 50 \\
\midrule
\multicolumn{5}{l}{\textbf{Measurement \& Performance Analysis (IF 8.49)}} & \multicolumn{5}{l}{\textbf{Operating Systems (IF 14.20)}} \\
IMC & 9.44 & 6 & 24.25\% & 108 & OSDI   & 18.88 & 14 & 18.20\% & 100 \\
SIGMETRICS       & 7.4 & 5 & 16.03\% & 93 & FAST & 14.06 & 10 & 19.25\% & 51 \\
\multicolumn{5}{l}{ }   & USENIX ATC & 13.91 & 7 & 20.85\% & 143 \\
\multicolumn{5}{l}{ }   & EuroSys & 12.78 & 7 & 19.69\% & 81 \\
\multicolumn{5}{l}{ }   & SOSP & 8.57 & 8 & 15.50\% & 54 \\
\midrule
\multicolumn{5}{l}{\textbf{Programming Languages (IF 8.18)}} & \multicolumn{5}{l}{\textbf{Software Engineering (IF 9.86)}} \\
PLDI & 10.86 & 7 & 22.55\% & 269 & ISSTA & 14.06 & 10 & 19.25\% & 51 \\
POPL       & 7.54 & 6 & 25.57\% & 129 & ICSE   & 13.28 & 9 & 21.65\% & 267 \\
ICFP & 6.18 & 4 & 31.90\% & 72 & FSE   & 8.1 & 6 & 26.25\% & 215 \\
OOPSLA       & 5.55 & 4 & 35.50\% & 187 & ASE & 6.67 & 4 & 18.27\% & 175 \\
\bottomrule
\end{tabular}%
}
\label{tab:addlabel}%
\end{table}%

\begin{table}
\caption{Scientometric overview of the field of Theory, grouped by research areas, tracking impact factors, median number of citations per article, acceptance rates, and number of accepted articles $N$ per conference.}
\makebox[\textwidth][c]{
\begin{tabular}{lcccr|lcccr}
\toprule
\multicolumn{10}{l}{Theory (IF 8.12)} \\
\midrule
Conf.\ & IF & median  & Acc.\ rate & $N$  & Conf.\ & IF & median  & Acc.\ rate & $N$ \\ \hline
\multicolumn{5}{l}{\textbf{Algorithms \& Complexity (IF 8.39)}} & \multicolumn{5}{l}{\textbf{Cryptography (IF 11.14)}} \\
STOC & 8.55 & 6 & 26.91\% & 247 & EuroCrypt & 11.54 & 7 & 20.55\% & 159 \\
SODA & 8.34 & 4 & 29.35\% & 361 & CRYPTO & 10.02 & 7 & 23.43\% & 56 \\
FOCS & 8.3 & 6 & 33.1\% & 244            & \multicolumn{5}{l}{ } \\
\midrule
\multicolumn{5}{l}{\textbf{Logic \& Verification (IF 5.30)}} & \multicolumn{5}{l}{} \\
CAV    & 6.9 & 4 & 27.05\% & 144 & \multicolumn{5}{l}{ } \\
LICS    & 3.93 & 3 & 39.90\% & 169 & \multicolumn{5}{l}{ } \\
\bottomrule
\end{tabular}%
}
\label{tab:addlabel}%
\end{table}%

\begin{table}
\caption{Scientometric overview of Interdisciplinary conferences, grouped by research areas, tracking impact factors, median number of citations per article, acceptance rates, and number of accepted articles $N$ per conference.}
\makebox[\textwidth][c]{
\begin{tabular}{lcccr|lcccr}
\toprule
\multicolumn{10}{l}{Interdisciplinary (IF 7.82)} \\
\midrule
Conf.\ & IF & median  & Acc.\ rate & $N$  & Conf.\ & IF & median  & Acc.\ rate & $N$ \\ \hline
\multicolumn{5}{l}{\textbf{Computational Biology \& Bioinformatics (IF 9.84)}} & \multicolumn{5}{l}{\textbf{Computer Graphics (IF 11.86)}} \\
RECOMB & 20.31 & 10 & 18.00\% & 297 & SIGGRAPH & 15.2 & 9 & 31.00\% & 272 \\
ISMB & 6.81 & 4 & 19.50\% & 120 & SIGGRAPH Asia & 12.0 & 7 & 35.00\% & 211 \\
\multicolumn{5}{l}{ } & EuroGraphics & 4.46 & 2 & 36.00\% & 127\\
\midrule
\multicolumn{5}{l}{\textbf{Computer Science Education (IF 4.50)}} & \multicolumn{5}{l}{\textbf{Economics \& Computation (IF 5.97)}} \\
SIGCSE    & 4.5 & 3 & 31.0\% & 341 & EC & 7.32 & 5 & 23.08\% & 139 \\
\multicolumn{5}{l}{ } & WINE & 2.83 & 1 & 30.515\% & 60 \\
\midrule
\multicolumn{5}{l}{\textbf{Human-Computer Interaction (IF 10.01)}} & \multicolumn{5}{l}{\textbf{Robotics (IF 6.74)}} \\
CHI    & 10.59 & 7 & 25.30\% & 1504 & RSS & 12.1 & 7 & 29.50\% & 191 \\
UIST    & 6.79 & 5 & 24.25\% & 192 & ICRA & 7.63 & 4 & 46.50\% & 3459 \\
UbiComp    & 4.29 & 3 & ?\% & 45 & IROS & 5.27 & 3 & 46.00\% & 2796 \\
\midrule
\multicolumn{5}{l}{\textbf{Visualization (IF 6.30)}} & \multicolumn{5}{l}{} \\
VIS    & 7.03 & 5 & 23.23\% & 421 & \multicolumn{5}{l}{ } \\
VR    & 5.53 & 4 & 22.20\% & 198 & \multicolumn{5}{l}{ } \\
ISMAR    & 5.08 & 4 & 21.24\% & 125 & \multicolumn{5}{l}{ } \\
\bottomrule
\end{tabular}%
}
\label{tab:addlabel}%
\end{table}%

We noted a broad spread of impact factors among the four research fields. In addition to the conference-specific numbers, the tables also report micro averages of citations per paper within each of the fields. AI (23.88 citations per paper on average) was significantly higher scored than Systems (10.27), Theory (8.12), and interdisciplinary outlets (7.82). We assume that this difference is a function of citation practices, general popularity or activity of the respective fields or venue selectiveness.

To follow up on this notion, we computed the Spearman Rank correlation between conference acceptance rates and impact factors, yielding a weak negative correlation of -0.27 ($p << 0.01$). This somewhat intuitive finding indicates that more selective conferences tend to show slightly higher impact. A much stronger effect was noticed when correlating the average impact factor per research area with the average number of papers accepted in that area. The resulting moderate correlation of 0.45 ($p < 0.05$) suggests that the overall popularity of a research area is a robust predictor of venue impact. 



\section{Discussion}
Love them or hate them, Impact Factors are one of the main ways in which many schools and departments measure scientific output and impact. Those of our colleagues working in non-CS departments have to regularly explain the selectiveness and import of CS conference proceedings to their journal-based colleagues. This broad empirical overview aims to give them additional data points, enabling a direct comparison to established journals. Most notably, top-ranked CS conferences such as ICLR (IF 48.87) or CVPR (IF 44.08) receive comparable impact factors to high-ranking journals.

There are several potential limitations to this investigation that should be kept in mind when interpreting our findings. (1) While publishers calculate impact factors using well curated catalogs, the numbers presented here are based on Google Scholar citations counts and therefore include incoming citations from non peer-reviewed sources such as preprint articles. The resulting impact factors are therefore potentially higher than what a reviewed-only experiment would yield. (2) For reasons of scope, we collected only incoming citation counts, not full citation sources that would allow tracking exact citation dynamics of what types of venues cite what type of publications. (3) At the time of writing this article, we are in the 2023 calendar year, making 2022 impact factors the latest scores available for reporting. As a consequence, however, we are studying publications from 2020 and 2021, the years most heavily affected by the COVID-19 pandemic. It is therefore possible that the citation patterns and impact factors reported for these years are not representative of the trends prior to, or after the pandemic.

\bibliographystyle{ACM-Reference-Format}
\bibliography{main.bib}


\begin{thebibliography}{3}


\ifx \showCODEN    \undefined \def \showCODEN     #1{\unskip}     \fi
\ifx \showDOI      \undefined \def \showDOI       #1{#1}\fi
\ifx \showISBNx    \undefined \def \showISBNx     #1{\unskip}     \fi
\ifx \showISBNxiii \undefined \def \showISBNxiii  #1{\unskip}     \fi
\ifx \showISSN     \undefined \def \showISSN      #1{\unskip}     \fi
\ifx \showLCCN     \undefined \def \showLCCN      #1{\unskip}     \fi
\ifx \shownote     \undefined \def \shownote      #1{#1}          \fi
\ifx \showarticletitle \undefined \def \showarticletitle #1{#1}   \fi
\ifx \showURL      \undefined \def \showURL       {\relax}        \fi
\providecommand\bibfield[2]{#2}
\providecommand\bibinfo[2]{#2}
\providecommand\natexlab[1]{#1}
\providecommand\showeprint[2][]{arXiv:#2}

\bibitem[Callaway et~al\mbox{.}(2016)]%
        {callaway2016beat}
\bibfield{author}{\bibinfo{person}{Ewen Callaway} {et~al\mbox{.}}} \bibinfo{year}{2016}\natexlab{}.
\newblock \showarticletitle{Beat it, impact factor! Publishing elite turns against controversial metric}.
\newblock \bibinfo{journal}{\emph{Nature}} \bibinfo{volume}{535}, \bibinfo{number}{7611} (\bibinfo{year}{2016}), \bibinfo{pages}{210--211}.
\newblock


\bibitem[Hutchins et~al\mbox{.}(2016)]%
        {hutchins2016relative}
\bibfield{author}{\bibinfo{person}{B~Ian Hutchins}, \bibinfo{person}{Xin Yuan}, \bibinfo{person}{James~M Anderson}, {and} \bibinfo{person}{George~M Santangelo}.} \bibinfo{year}{2016}\natexlab{}.
\newblock \showarticletitle{Relative citation ratio (RCR): a new metric that uses citation rates to measure influence at the article level}.
\newblock \bibinfo{journal}{\emph{PLoS biology}} \bibinfo{volume}{14}, \bibinfo{number}{9} (\bibinfo{year}{2016}), \bibinfo{pages}{e1002541}.
\newblock


\bibitem[of~Science Editors~(EASE)(2007)]%
        {european2007ease}
\bibfield{author}{\bibinfo{person}{European~Association of Science Editors~(EASE)}.} \bibinfo{year}{2007}\natexlab{}.
\newblock \showarticletitle{EASE statement on inappropriate use of impact factors}.
\newblock \bibinfo{journal}{\emph{Eur Sci Edit}} \bibinfo{volume}{33}, \bibinfo{number}{4} (\bibinfo{year}{2007}), \bibinfo{pages}{99}.
\newblock


\end{thebibliography}

\end{document}